\documentclass[conference]{IEEEtran}

\usepackage{graphicx}
\usepackage{epstopdf}
\usepackage{url}
\usepackage{amsmath,amssymb,amsfonts}
\usepackage{caption,subcaption}
\usepackage{comment}
\usepackage{marginnote}
\usepackage{pifont}
\usepackage{amsmath,amsfonts,amssymb}
\usepackage[draft]{minted}
\usepackage{caption}
\usepackage{subcaption}
\usepackage[linesnumbered,ruled,vlined]{algorithm2e}
\usepackage[export]{adjustbox}
\usepackage{listings}
\usepackage{enumitem}
\usepackage{tabularx}
\usepackage{mathtools}
\usepackage{nccmath}
\usepackage{hhline}
\usepackage{makecell}
\usepackage{bm}

\newcommand{\distressnetNG}{Anonymous~}

\DeclarePairedDelimiter{\ceil}{\lceil}{\rceil}
\newcommand{\xmark}{\ding{53}}

\begin{document}

\title{R-Drive: Resilient Data Storage and Sharing for Mobile Edge Computing Systems}


\author{
M. Sagor, R. Stoleru, S. Bhunia, M. Chao, A. Haroon, A. Altaweel, M. Maurice$^{\dagger}$, R. Blalock$^{\dagger}$\\
    \IEEEauthorblockA{Department of Computer Science and Engineering, Texas A\&M University\\
    $^{\dagger}$National Institute of Standards and Technology (NIST)}
    \IEEEauthorblockA{[msagor, stoleru, sbhunia, chaomengyuan, amran.haroon, altaweelala1983]@tamu.edu}
}

\maketitle





\newcommand{\schname}{\textit{R-Drive}}
\begin{abstract}
Mobile edge computing (MEC) systems (in which intensive computation and data storage tasks are performed locally, due to absence of communication infrastructure for connectivity to cloud) are currently being developed for disaster response applications and for tactical environments. MEC applications for these scenarios generate and process significant mission critical and personal data that require resilient and secure storage and sharing. In this paper we present the design, implementation and evaluation of R-Drive, a \textit{resilient} data storage and sharing framework for disaster response and tactical MEC applications. R-Drive employs erasure coding and data encryption, ensuring resilient and secure data storage against device failure. R-Drive adaptively chooses erasure coding parameters to ensure highest data availability with minimal storage cost. R-Drive's distributed directory service provides a resilient and secure namespace for files with rigorous access control management. R-Drive leverages opportunistic networking, allowing data storage and sharing in mobile and loosely connected edge computing environments. We implemented R-Drive on Android, integrated it with existing MEC applications. Performance evaluation results show that R-Drive enables resilient and secure data storage and sharing.
\end{abstract}
\section{Introduction}
\begin{figure}[t]
\centering
\includegraphics[width=0.85\linewidth]{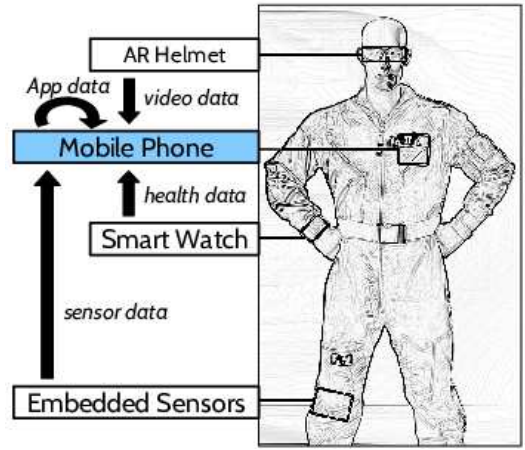}
\caption{Next generation first responders equipped with wearable technologies including AR helmets, mobile phones, smart watches and embedded sensors, which generate large amounts of data that require resilient storage for processing~\cite{nextGenerationFR} }
\label{fig:first_responder}
\end{figure}

\vspace{10pt}
Mobile Edge Computing (MEC) has gained significant popularity over traditional cloud computing due to low latency guarantee for data storage and processing. In this architecture, devices form a local cloud using available computing and storage resources, allowing applications to process and store data locally~\cite{ahmed2017mobile, premsankar2018edge, sun2016edgeiot, namburu2020advances, ray2017framework}. Edge computing platforms that are designed for mobility need to handle disconnected environments where infrastructure networks such as cellular and Wi-Fi are unavailable and cloud services are unreachable~\cite{olaniyan2018opportunistic, cui2017software, lu2017cooperative}. In such cases, MEC applications are entirely dependent on available edge resources for operations. Disconnection-tolerant MEC platforms for disaster response and wide area search and rescue operations are gaining significant popularity~\cite{boukerche2018smart, chenji2014fog, reina2015survey}. In such scenarios, first responders are equipped with necessary hardware including a manpack, mobile devices, wearable gadgets, sensors etc., to perform mission critical operations (Figure~\ref{fig:first_responder}).


Devices employed in MEC, e.g., on-body cameras, smart watches, gesture recognition devices, body sensors (heat, gas, water etc.) as well as MEC applications on mobile devices generate large amounts of mission critical data and perform storage intensive tasks. Storage intensive tasks such as urban sensing, survey collection, geo-spatial data collection, text and media files storage and any other quantitative and qualitative data storage by users etc., require \textit{resilient data storage} with low overhead~\cite{calabrese2014urban, rahman2017towards}. Device failure can occur due to hardware malfunction and battery depletion due to heavy use of edge devices. Hence, data storage in mobile edge must employ replication based distributed storage so that data is not lost due to device failure. Additionally, \textit{disconnection resilient data sharing} among entities in MEC is difficult due to absence of infrastructure networks and frequent device mobility. In a search and rescue scenario, since first responders perform their respective tasks being agnostic of network connectivity, data sharing among entities needs to be network disconnection tolerant. 

Existing file/data storage services, e.g., Dropbox~\cite{Dropbox}, Google Drive~\cite{googledrive}, OneDrive~\cite{onedrive} etc. are not designed for MEC and cannot operate in the absence of cloud. Although these services allow users to store and modify files offline, the files are simply stored locally, making them prone to data unavailability/loss due to device failure by energy depletion or hardware malfunction. Moreover, mobile devices at the edge are prone to frequent disconnection/separation from the network, due to mobility or network congestion. Thus, local updates may not propagate to the cloud despite intermittent cloud access. Thus, there is a need for a cloud-like data storage service at the edge that uses available edge resources for storing data.

Existing storage services also enable \textit{data sharing} among devices. This can only be accomplished through the cloud via infrastructure networks. Users may employ data sharing applications that make use of ad-hoc network connectivity (e.g., Bluetooth, Wi-Fi Direct). But, disconnection may occur during a data sharing session; thus, users are required to minimize movement and stay connected until the data sharing session completes. Hence, traditional data sharing schemes are impractical for disconnection oriented mobile edge.

To address the aforementioned limitations, we designed and implemented \schname, a data storage and sharing framework for MEC environments. \schname\ handles both device and network failures in MEC environments, eliminating the above mentioned data storage and sharing problems by bringing cloud services to the edge. \schname\ utilizes available storage resources of devices to the edge to resiliently store data and allows users/applications to securely share them with proper access control. The key features of \schname\ and the contributions of the paper are as follows:


\begin{itemize}
\item \schname\ employs distributed data storage with encryption and erasure coding, enabling resilient data storage against device failure.
\item \schname\ employs opportunistic networking, maximizing the use of available bandwidth, at the same time abstracting network failure from client applications.
\item \schname\ incorporates resilient distributed directory service with secure access control.
\item \schname\ transparently enables existing data storage applications to share data without assistance from the cloud.
\end{itemize}



\section{Background and State of the Art}
\label{sec:background}



\subsection{Mobile Edge Computing for Disaster Response and Tactical Environments}

\begin{figure}[t]
\centering
\includegraphics[width=0.90\linewidth]{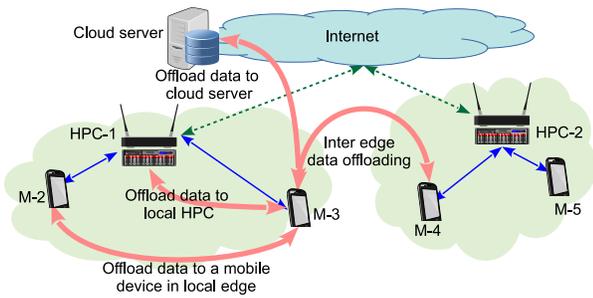}
\caption{Mobile devices form two mobile edge networks (MEC-1 and MEC-2) and share resources among themselves, or with the cloud, to perform data storage/sharing and processing tasks.}
\label{fig:arch_edge}
\end{figure}

Figure~\ref{fig:arch_edge} depicts a MEC architecture for disaster response or tactical environments. Multiple mobile devices form an edge network that can be disconnected from the Internet and cloud servers. Mobile devices may be connected to a \textit{HPC node} that manages communications (e.g., LTE, Wi-Fi, Wi-Fi Direct), IP addresses allocations to devices, and DNS services, mapping device names to their corresponding IPs. The central node also performs high-performance computing (HPC) functions on data produced by edge devices. In addition to the HPC node, data can also be offloaded to connected mobile devices for processing and storage. As shown in the figure, two edge networks (where nodes HPC-1 and M-6 serve as central nodes for edges 1 and 2, respectively) can be connected over Internet, or they can discover each other locally.

\begin{figure}[t]
\centering
\includegraphics[width=0.85\linewidth]{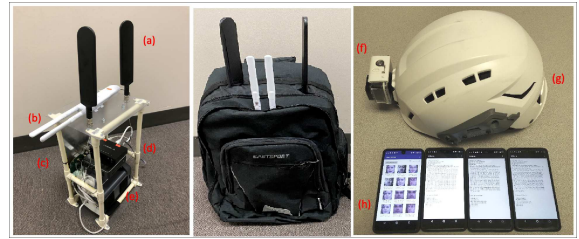}
\caption{\distressnetNG - a MEC platform for disaster response consisting of: a) LTE antenna, b) Wi-Fi AP, c) LTE eNB, d) Intel NUC that runs LTE EPC and HPC; e) Battery; f-g) Helmet with body camera; h) Android phones}
\label{fig:distressnet-ng-hardware}
\end{figure}

\begin{figure}[t]
\centering
\includegraphics[width=0.85\linewidth]{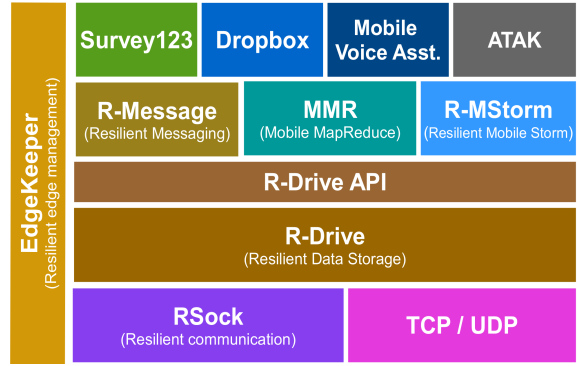}
\caption{Software ecosystem for the \distressnetNG mobile edge computing suit}
\label{fig:distressnet-ecosystem}
\end{figure}

An example of a MEC system for disaster response is shown in Figure~\ref{fig:distressnet-ng-hardware}. The system \distressnetNG, consists of a manpack equipped with wireless communication (LTE and Wi-Fi) and processing capabilities (HPC). Multiple mobile devices communicate among themselves and with the manpack using wireless communication, for sharing data storage and computational resources. The software architecture for \distressnetNG is shown in Figure~\ref{fig:distressnet-ecosystem}. As shown, several applications have been specifically designed for MEC. For example, R-MStorm (Resilient Mobile Storm)~\cite{chao2020r} is for real-time stream processing, MMR (Mobile Map Reduce) for batch processing, a mobile virtual voice assistant for emergency medical services, edge resource orchestration, etc. Survey123~\cite{survey123} is used by several disaster response teams (e.g., Texas Task Force) to gather field data (e.g., number of survivors, hazardous locations etc.) and send it to a database located either in the cloud or a local server. Survey123 can operate in completely disconnected environments, however, in such environments, the data is cached locally on mobile devices and only uploaded once a cellular or Wi-Fi connection to the Internet is established. ATAK~\cite{ATAK} is an Android application used by the US military to share mission critical data during combat missions or disaster response operations. Similar to Survey123, ATAK stores data on local storage if communication with a master ATAK server is unavailable.

As shown in Figure~\ref{fig:distressnet-ecosystem}, two important components of the \distressnetNG~architecture are EdgeKeeper and RSock (Resilient Socket). \textbf{EdgeKeeper}~\cite{bhuniaedgekeeper} is a distributed coordination and service discovery, and meta-data storage application which runs on all devices of the edge. EdgeKeeper is based on a primary/master-replica/slave architecture in which at one time one device acts as a master, whereas other devices act as slaves. EdgeKeeper master is responsible for maintaining distributed consensus among devices and providing services such as device authentication, service discovery, edge health monitoring, network topology management, metadata store etc. EdgeKeeper slaves are standby devices to take over dead master and maintain services. EdgeKeeper employs \textit{Globally Unique Identifier} (GUID)~\cite{sharma2014global} to uniquely identify each edge device. Each GUID is a unique 40 characters long string, generated with a unique public and private key pair, assigned to one user. EdgeKeeper is responsible for performing DNS-like GUID to/from IP mapping. GUID based device identification allows applications on different devices to communicate with each other being agnostic of IP assignments. Such identification scheme enables mobile devices pertaining to different edges to communicate, and perform inter-edge tasks. \textbf{RSock}~\cite{altaweel2020rsock} is a resilient transport protocol designed for sparsely connected network environments aiming to make best use of available network bandwidth and to ensure timely data delivery. RSock provides routing by GUID and replication of packets, to be sent over multiple paths for device-to-device communication (i.e., using any available wireless interface - LTE, Wi-Fi, Wi-Fi Direct). RSock employs the Hybrid Routing Protocol (HRP)~\cite{yang2016hybrid}, which performs packet replication to reduce the packet delivery delay. A RSock header contains a sequence number for the corresponding packet so that receiver can assemble the packet in its respective position. For a file to be received successfully, all RSock packets must have to be received and assembled by their packet numbers at the receiver. RSock API allows user to set a time to live (TTL) value for a packet in the header. The TTL value entails for how long (in seconds) a packet can be alive in the network. If the TTL for a packet expires, the packet will be immediately discarded from the network.


\subsection{Motivation and State of the Art}
\vspace{10pt}
Applications in MEC platforms for disaster response generate significant amounts (e.g., gigabytes) of mission critical and personal data that require resilient and secure storage. As examples, R-MStorm generates media data about disaster victims, Mobile Voice Assistant collects patients' personal medical information, Survey123 collects various survey data during search and rescue operations. Currently, this data is stored only on a device's local storage; if a device's storage runs out, these applications cannot store new data. Also, if the device fails, the data stored on it is lost or inaccessible. Existing data storage services such as Dropbox, Google Drive, OneDrive etc. \textit{require cloud connectivity to store data.} During large scale disasters, infrastructure networks such as LTE, Wi-Fi, etc. may not be available, hence above services can not be used for storing data. Moreover, device failure may occur due to energy depletion or hardware malfunction. Hence, \textit{data is vulnerable to loss/unavailability} if stored on a single device without added resilience. Pure data replication across devices to ensure resilience against device failure is not a feasible solution for mobile edge due to limited storage availability. Furthermore, some storage services store offline files in local storage without any protection (i.e., encryption), allowing \textit{data tampering by injecting corrupt data} by malicious applications.

We conducted experiments with above mentioned storage services and observed that these services do not provide both data resilience and security when storing locally. We stored large amount of offline data and observed that, when device storage runs out, these services become inoperational, despite other devices in same network having large amounts of available storage. We conducted another set of experiments in which we tampered offline data of Dropbox and Survey123 by injecting malicious data and observed that corrupted updates were later propagated to cloud when applications were started. Google Drive and OneDrive do not store offline files in external storage, hence they are not directly accessible via Android filesystem. But, the offline files are still vulnerable of device failure as they are stored in single device. Table~\ref{tab:table_motivation_1} summarizes the limitations of existing storage solutions, as well as storage intensive MEC applications.


\begin{table}[ht]
        \small
		\centering
		\scalebox{1.0}{
		\begin{tabular}{l|c|c|c} 
		    \textbf{} & \textbf{Cloudless} &\textbf{Resilient} & \textbf{Offline} \\
		    \textbf{Services} & \textbf{Storage} & \textbf{Storage} & \textbf{Encrypt} \\
            \hline
            Dropbox & \xmark &  \xmark & \xmark \\
            \hline
            OneDrive & \xmark &  \xmark & \checkmark \\
            \hline
            Google Drive & \xmark &  \xmark & \checkmark \\
            \hline
            Survey123 & \xmark &  \xmark & \xmark \\
            \hline
            R-MStorm & \checkmark &  \xmark & \xmark \\
            \hline
            \textbf{R-Drive} & \pmb{\checkmark} & \pmb{\checkmark} & \pmb{\checkmark}
        \end{tabular}
        }
		\caption{Existing storage services and MEC applications rely on cloud connectivity for data storage.}
		\label{tab:table_motivation_1}
\end{table}

To further address the limitations, we investigated state of the art distributed storage and file system solutions for mobile edge. CODA~\cite{kistler1992disconnected}, maintains a local cache during disconnected operations to store edited data and requires cloud connectivity to synchronize local cache with replica servers. OFS~\cite{paiker2017design}, carries heavy storage overhead since it keeps a copy of the same data to both local device and cloud to ensure data availability. MEFS~\cite{scotece2019mefs} tried to retro-fit a cloud based file system to use for mobile edge, but the solution greatly relied on cloud communication. PFS~\cite{dwyer1997mobility}, FogFS~\cite{pamboris2019fogfs} rely in specific mobility models that makes them impractical for disconnected mobile edge. Although HDFS~\cite{shvachko2010hadoop} and GFS~\cite{ghemawat2003google} use erasure coding instead of replication to store data in replica devices, these solutions are too heavy-weight for mobile devices in terms of memory footprint and computation. Hyrax~\cite{marinelli2009hyrax} tried to port HDFS for Android devices and experimented in mesh networks. Despite decent engineering efforts, Hyrax showed poor performance for CPU bound tasks. MDFS~\cite{DBLP:journals/tcc/ChenWSX15} implementation was based on a purely connected network, which is a major fallacy in disconnected mobile edges. MDFS did not provide a file system-like functionality, such as directory service, access control management etc.


\begin{table}[ht]
        \small
		\centering
		\scalebox{1.0}{
		\begin{tabular}{l|c|c|c}
		    \textbf{} & \textbf{Cloudless}  & \textbf{Opportunistic} & \textbf{Cloudless}\\ 
		    \textbf{Services} &\textbf{Share} & \textbf{Share} & \textbf{Namespace}\\ 
		    \hline
		    Dropbox & \xmark & \xmark & \xmark \\
            \hline
            OneDrive & \xmark & \xmark & \xmark \\
            \hline
            Google Drive & \xmark & \xmark & \xmark \\
            \hline
            Share Apps & \xmark & \xmark & \xmark \\
            \hline
            \textbf{R-Drive} & \pmb{\checkmark} & \pmb{\checkmark} & \pmb{\checkmark} 
        \end{tabular}
        }
		\caption{Existing data sharing services cannot share data without cloud connectivity.}
		\label{tab:table_motivation_2}
\end{table}

As mentioned earlier, infrastructure networks may be unavailable during large scale disasters. Existing services such as Dropbox, Google Drive, OneDrive etc. can only \textit{share data across devices if there is connectivity to the cloud.} Users can use file sharing applications (Google Files~\cite{googlefiles}, SHAREit~\cite{SHAREit} etc.) that do not require cloud connectivity and can operate over ad-hoc networks such as Wi-Fi Direct, Bluetooth, NFC etc. But, ad-hoc networks rely on short range communication and constant connectivity. During large scale search and rescue operations, team members are mobile and scattered across large areas; it may not always be possible for two team members to stay within each other's communication range to exchange data. For instance, two team members may not be directly connected yet reachable via one or many intermediate mediums/people that frequently travel back and forth between them. Consequently, MEC platforms for disaster response should support \textit{network opportunistic data sharing over multiple hops.} Moreover, during disaster response operations, first responders are divided into groups to perform their respective tasks. Team members often need to share mission critical data among themselves to co-ordinate their tasks. Also, sharing data with other teams require rigorous access control so that critical data is only shared with authorized personnel (e.g., team leaders). Existing data sharing services cannot sync directories across devices in absence of cloud connectivity. Consequently, first responder teams need to have a \textit{common namespace to manage data and permissions} that does not rely on cloud connectivity. Table~\ref{tab:table_motivation_2} summarizes the limitations of the existing data sharing schemes in disconnection oriented MEC platforms. 

To apply erasure coding for data storage, two important input parameters are $n$ and $k$. A high $n$ and low $k$ increases data availability at a cost of storage, and vice-versa. $(n,k)$ should be decided dynamically depending on the edge resource availability and user provided quality of service (QoS) parameters. Although HDFS~\cite{hdfsEC}, GFS~\cite{ghemawat2003google} use erasure coding for distributed storage in a cluster, the choice of parameters for erasure coding $(n,k)$ is fixed, since both HDFS and GFS were designed primarily for large storage clusters consisting of hundreds of storage nodes residing in stationary racks, in a data-center. MDFS~\cite{DBLP:journals/tcc/ChenWSX15} incorporated erasure coding for disconnection prone mobile edge, and took network link quality, energy cost etc. in consideration. But they did not provide any online algorithm to select $n$ and $k$ values for variable storage availability and file sizes. Zhu et al., presented an online adaptive code rate selection algorithm for cloud storage~\cite{zhuonline}. The algorithm takes real-time user demands as one of the input metrics to a regret minimization problems to decide the most optimum $n$ and $k$ values. The solution states as one of their assumptions is that all the candidate storage devices have enough storage capabilities, which can be a big assumption for mobile edge computing environment where devices are heterogeneous. HACFS~\cite{xia2015tale} is another novel solution, where authors implemented an extension to HDFS to adaptively choose between fast code and compact code depending on data read hotness. The solution also up-codes and down-codes previously encoded data to ensure data resiliency against loss due to various reasons. Despite having the capability of switching between two coding schemes, their solution involves using fixed coding parameters for each of the coding schemes. Zhang et al., proposed an erasure coded storage system consisting Android devices, also provided no study for choosing the most efficient nodes and $n$, $k$ values~\cite{zhang2020design}. Shu et al., proposed an erasure coded distributed storage system on fog nodes, but provided no analysis as to how to choose the $n$, $k$ values~\cite{shu2018binary}.



\section{R-Drive System}
\label{sec:design}

\begin{figure}[t]
	\centering
	\includegraphics[width=0.85\linewidth]{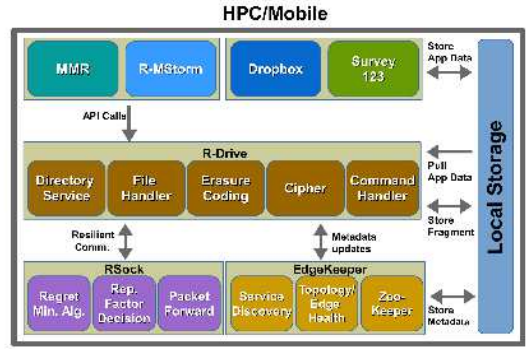}
	\caption{\schname\: components and integration with the \distressnetNG~software ecosystem}
	\label{fig:r-drive_architecture}
\end{figure}


\subsection{R-Drive System Overview}
\subsubsection{System Components}
Figure~\ref{fig:r-drive_architecture} shows the \schname\ system components and its integration with the \distressnetNG~software ecosystem. As shown, \schname\ consists of five major components. \textbf{Directory Service} provides a namespace for all files and directories. \textbf{File Handler} performs file and directory operations (e.g., creation, retrieval and removal).  \textbf{Erasure Coding} component encodes and decodes data into fragments using Reed-Solomon erasure coding~\cite{reed1960polynomial}. \textbf{Cipher} encrypts and decrypts data using 256 bit AES encryption. \textbf{Command Handler} handles commands for basic storage operations such as \textit{-put, -get, -mkdir, -ls, -rm} etc.

Client applications such as MMR, R-MStorm use the \schname\ API to perform data storage and sharing operations. Applications such as Dropbox, Survey123 etc. generate and store \textit{app data} on device's local storage. \schname\ periodically fetches the \textit{app data} and stores it in \schname\ for resilience against device failure. \schname\ communicates with RSock and EdgeKeeper applications via JSON based RPC calls over local TCP sockets.


\subsubsection{Data storage in R-Drive}
Storage in \schname\ takes place via \schname\ user interface (UI) or Java client API. Client applications can make appropriate \schname\ API calls to perform storage and sharing tasks. \schname\ UI allows a device operator to directly interact with the application. \schname\ also provides command line interface (CLI) that allows a desktop user (e.g., a search and rescue operation coordinator) to connect to a device in the field and perform \schname\ operations with permissions. The \schname\ API is as follows:

\begin{minted}{java}
int mkdir(String rdriveDirectory, 
            List<String> permissionList);
List<String> ls(String rdriveDirectory);
int put(String localFilePath, 
            String rdriveFilePath,
            List<String> permissionList);
int get(String rdriveFilePath, 
            String localFilePath); 
int rm(String rdriveFilePath);           
\end{minted}

\schname\ also allows data storage by monitoring files in user-selected directories on local storage, similar to Google Backup and Sync~\cite{backupandsync}. A user selects a directory in the local storage that \schname\ will monitor for any changes in data and periodically pulls the data to store in \schname\ system. User can select application directories which are prone to data loss due to device failure. Currently, \schname\ supports backing up \textit{app data} for Survey123, ATAK, Dropbox.

\subsubsection{Data sharing in R-Drive}
\schname\ enables inter device data sharing using RSock communication channel. A device can share data to one or multiple devices as follows: a) unicast the data to any device; b) upload the data to \schname\ system and set permission appropriately for authorized users. In scenario (a), no data erasure coding takes place and the entire file is sent. The TTL value in RSock is set appropriately, to decrease the likelihood of data being discarded during routing, but to also not congest the network (by keeping too many copies of the data).

\subsection{Directory Service}
\label{subsec:ds}

The Directory Service provides all metadata-related operations such as creation of new metadata, retrieval of existing metadata, checking metadata permissions, presenting a namespace to clients etc. \schname\ maintains a hierarchical directory structure; the top level directory is root(/) and below are subsequent subdirectories. A metadata in \schname\ system is called an \textbf{rnode}, with its structure shown in Table~\ref{tab:table_rnode}). A rnode represents either a file or a directory entity in \schname. After creating a rnode, the Directory Service stores it in EdgeKeeper, which internally stores it in ZooKeeper data nodes, commonly known as znode. ZooKeeper replicates znodes to replica devices for fault tolerance, to handle master failure or cluster disconnection. If one or more replica devices leave the edge, other edge devices become new replicas and the lost znodes are replicated to the new replica devices. Consequently, if EdgeKeeper has \textit{r} replicas, then \schname\ metadata remains intact and it can provide Directory Services despite device failures as long as there are $\ceil*{r/2}$ devices available at the edge. 

A directory creation takes place when a client invokes the \textit{mkdir()} API function or when the command \textit{-mkdir} is executed in the CLI. The Directory Service creates a new rnode for the target directory. Directory Service then fetches a copy of immediate parent rnode of the target directory from EdgeKeeper and inserts the target rnode information in the parent rnode. Finally, Directory Service pushes both parent and target rnodes to EdgeKeeper. EdgeKeeper stores the rnodes in ZooKeeper as data nodes. Directory retrieval is initiated when a client invokes the \textit{get()} API function or when the command \textit{-ls} is executed in the CLI.

\begin{table}
  \begin{center}
    \caption{\schname\ rnode structure}
    \label{tab:table_rnode}
    \scalebox{0.90}{
    \begin{tabular}{lcr} \hline
      \textbf{Field} & \textbf{Size} & \textbf{Description}\\
      \hline
      rnodeType    & 1 Byte   & File or Directory rnode\\
      rnodeID      & 16 Bytes & Unique rnode ID\\
      fileName     & Variable & Original File Name\\
      fileSize     & 8 Bytes & Original File Size\\
      fileID       & 16 Bytes & Unique File ID\\
      filePath     & Variable & R-Drive File Path\\
      N            & 2 Bytes  & N value for EC\\
      K            & 2 Bytes  & K value for EC\\
      blockCount   & 2 Bytes  & Number of Blocks\\
      fragLocation & Variable & locations of fragments\\
      fileList     & Variable & List of Files\\
      folderList   & Variable & List of Subdirectories\\
      permission   & Variable & Access Control List\\
      timeStamp    & 8 Bytes  & Time of Creation \\\hline
    \end{tabular}
    }
  \end{center}
\end{table}

\subsubsection{Access Control Management}
\schname\ leverages ZooKeeper's access control for managing permissions for rnodes. ZooKeeper~\cite{zkOfficial} supports pluggable authentication schemes. \schname\ implements its own custom authentication scheme as part of the Directory Service. \schname\ follows the standard UNIX permission scheme which can be set during file or directory creation. Permissions can also be set via the \textit{-setfacl} and \textit{-getfacl} commands entered through the CLI. A file or directory creator can set permission for any rnode for OWNER, WORLD, or a list of GUIDs. Each rnode permission only pertains to itself and does not apply to children. 

\subsection{R-Drive Data Storage}
\label{subsec:storage}

\begin{figure}[t]
	\centering
	\includegraphics[width=0.85\linewidth]{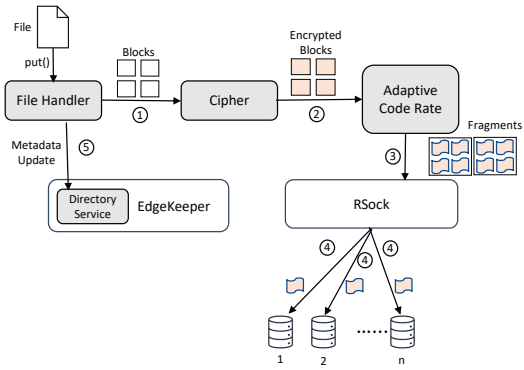}
	\caption{\schname\ file storage steps: partitioning the file into blocks, encrypting them, applying the adaptive erasure coding and distributing the fragments to best suitable $n$ nodes}
	\label{fig:distressnet-ng_fc}
\end{figure}

All data is stored in \schname\ as files. File creation involves copying a file from local file system to \schname\ using the \textit{put()} API function or the \textit{-put} command. The File Handler loads the target file and divides it into fixed sized blocks. Each block is then encrypted with a unique secret key and later converted into $n$ fragments using erasure coding. Directory Service communicates with EdgeKeeper to push the rnode for newly created file. If the rnode update succeeds, all fragments are sent through RSock by invoking the RSock client API. All fragments contain a timestamp that acts as a version number for fragments. A receiver device only accepts fragments with same or higher timestamps. Figure~\ref{fig:distressnet-ng_fc} shows the steps for a file creation process in \schname\ .

\subsubsection{R-Drive Data Encryption}
\schname\ uses 256 bit AES encryption using a unique secret key for file encryption. The key is further divided into $B$ key-shards using Shamir's Secret Sharing Scheme (SSSS)~\cite{shamir1979share}. SSSS is a distributed secret sharing scheme in which a secret is divided into shards in such a way that individual shards cannot reveal any part of the secret, whereas an allowed number of shards put together can reveal the secret. $(T,N)$ is the conventional way to express the SSSS system, where $N$ is the total number of secret shards, and $T$ is the minimum number of shards required to unveil the secret. In \schname\ we used $(B,B)$ as parameters for SSSS, where $B$ is the number of blocks. 

\subsubsection{Resilient Storage through Erasure Coding}
\schname\ uses Reed-Solomon erasure coding for data redundancy~\cite{reed1960polynomial}. Erasure codes are forward error correction (FEC) techniques that take a message of length $M$ and convert it into coded message of length greater than $M$ by adding redundancy so that the original message can be reconstructed by a subset of the coded message. In \schname\ storage, a file of size $F$ is divided into $k$ fragments, each of size $F/k$. Applying $(n,k)$ encoding on $k$ fragments will result in $n$ fragments, each of size $F/k$, where $n \geq k$. Hence, the total file size will be $n \cdot F/k$. Encoded $n$ fragments are then stored in geographically separated storage devices. To reconstruct the file, any $k$ fragments are sufficient. Thus, the system tolerates up to $n-k$ device failures. 

\subsubsection{Adaptive Code Rate}
The most widely used erasure coding library is Reed-Solomon that takes $(n, k)$ as parameters. The choice of $n$ and $k$ values are directly related to data redundancy (hence availability) at cost of additional storage overhead. So, choosing devices that has enough available storage is the basic requirement for storing data in mobile edge. Also, devices in edge are prone to device failure due to energy exhaustion, hardware failure, etc. So, choosing the devices that has more chance of survival against device failure is also an important factor to consider. Hence, in summary, the problem statement is, how to choose the best $n$ and $k$ values, and the fittest $n$ nodes (in terms of available battery life, storage capacity etc) so that the entire edge system can achieve highest data availability for the least storage cost.

\begin{figure}
	\centering
	\includegraphics[width=0.55\linewidth, angle=-90]{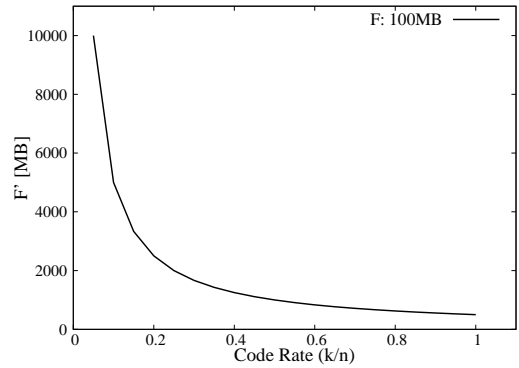}
	\caption{File size F' after erasure coding (applied to a file F of size 100MB) as a function of the code rate}
	\label{fig:cr_vs_fstar}
\end{figure}

The ratio $k/n$ in erasure coding, or the \textbf{\textit{code rate}}, indicates the proportion of data bits that are non-redundant. As a rule of thumb, when code rate decreases, the file size after erasure coding increases, and vice-versa. But, at the same time, lower code rate usually comes with a higher $n$ and lower $k$ values, providing added fault tolerance to the data. So, we cannot simply choose the lowest possible code rates; in that case, we will exhaust the system storage capacity very rapidly. Figure~\ref{fig:cr_vs_fstar} shows the file size after erasure coding as a function of code rate to illustrate the fact that erasure coded file size increases exponentially with decreasing code rate.

We need an online algorithm that dynamically chooses the $(n,k)$ values, and the fittest $n$ nodes for file storage in the edge. The algorithm's main focus will be to incorporate edge specific parameters (remaining battery, available storage, user/file specific quality of service parameters etc) to decide $n$ and $k$ values to optimize between data availability and storage cost. To reduce complexity, we will avoid all Reed-Solomon library specific parameters and use the default values for them.

To enable erasure coding in \schname, we need to answer the following: \textit{1) What code rate and what $(k,n)$ pair should the system choose? 2) Given a chosen code rate and $(k,n)$ value pair, which specific $n$ devices should the system store the $n$ file fragments to? 3) What system parameters used in answering 1) and 2) will be collected, and how?}

\noindent\textbf{Q1: What $k$ and $n$ values?} Code rate is calculated as $k/n$. If $k/n$ is small, there is a high probability to recover a file because only a small number of file fragments are required to reconstruct the original file. The file size after erasure coding with code rate $k/n$ is calculated as $F'=F*n/k$, where $F$ represents the original file size. In this case, if $k/n$ is too small, $n/k$ becomes very large, then the encrypted file size $F'$ becomes very large as well. Small $(k,n)$ entails higher file availability at a cost of larger storage overhead for the entire \schname\ system. To address this trade-off, we present the cost of availability and storage $C$ as a weighted sum and formulate the problem as a minimization problem as follows:

\begin{align}
\label{eq:minimization}
\underset{(k,n)} {\text{minimize}} \quad  & C(k,n,w_a) = w_a*k/n+(1-w_a)*n/k \\
\text{subject to:}  \quad  &  F/k \leq S_n,\\
                           &  T \leq T_k,  \\
                           &  1/N \leq k \leq n \leq N, k,n\in Z^+ \\
                           &  0 \leq w_a \leq 1
\end{align}

where $w_a$ denotes the weight of availability cost, $1-w_a$ the weight of storage cost, $S_n$ the $n^{th}$ maximum available storage of all nodes, $T_k$ denotes the $k^{th}$ longest remaining time among the total available $N$ devices, $T$ denotes the minimum time that a file is expected to be available in \schname. In the minimization problem, constraint~(2) ensures that the storage allocation for a node does not exceed available storage, constraint~(3) ensures that only devices with enough battery (for the selected lifetime $T$ of a file) are selected, constraint~(4) ensures that only positive $n$ and $k$ are selected, in the range $[1/N, N]$.

The weight $w_a$ is adjusted adaptively for different files; for a critical file, the system sets a large $w_a$ so that a small $k/n$ is chosen to improve its availability; for a large but unimportant file, the system sets a small $w_a$ so that a small $n/k$ is chosen to reduce the total storage cost.

\begin{figure}[ht]
    \centering
    \begin{subfigure}{0.32\linewidth}
        \includegraphics[width=\linewidth]{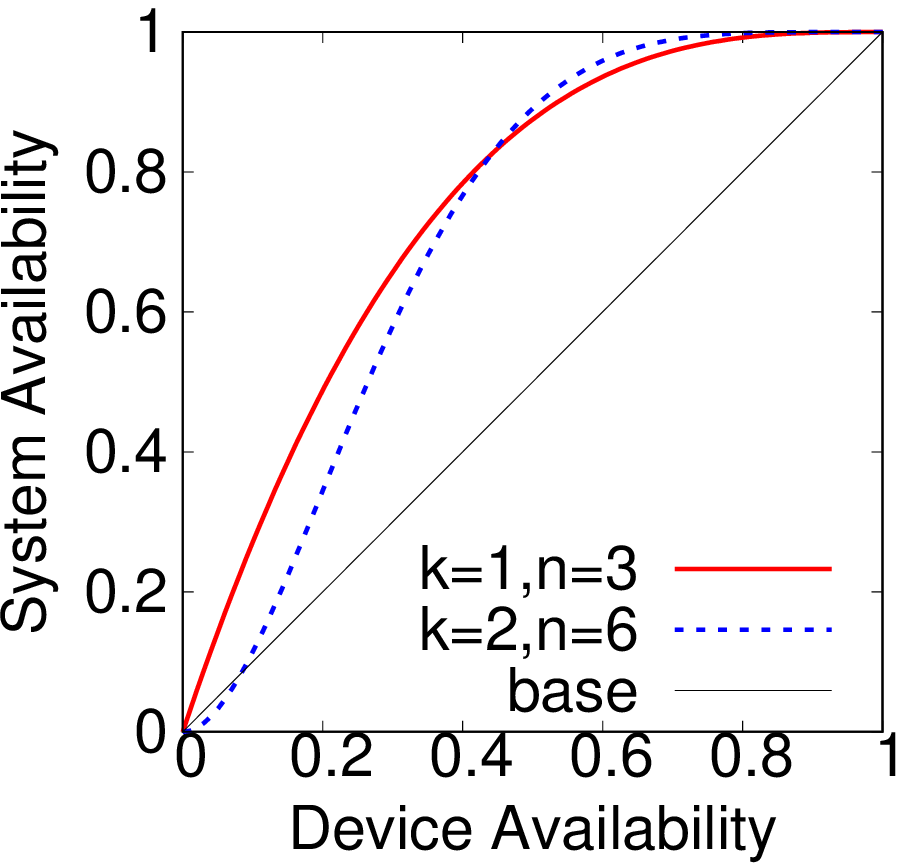}
        \subcaption{k/n=1/3}
        \label{fig:k1n3}
    \end{subfigure}
    \begin{subfigure}{0.32\linewidth}
        \includegraphics[width=\linewidth]{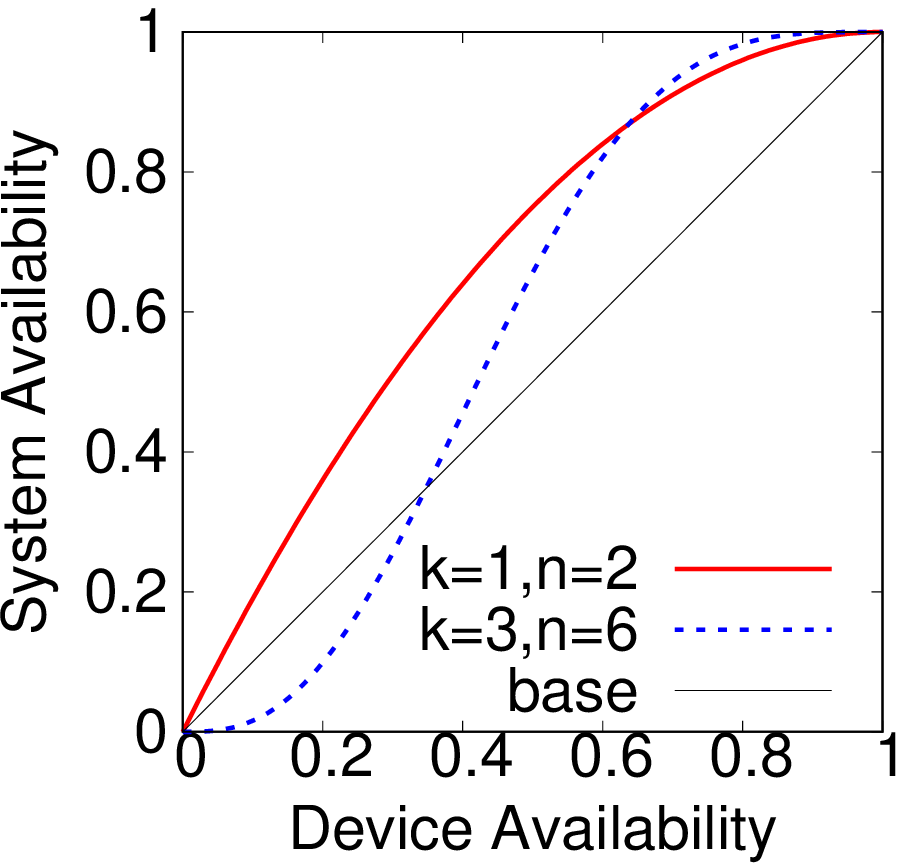}
        \subcaption{k/n=1/2}
        \label{fig:k1n2}
    \end{subfigure}
    \begin{subfigure}{0.32\linewidth}
        \includegraphics[width=\linewidth]{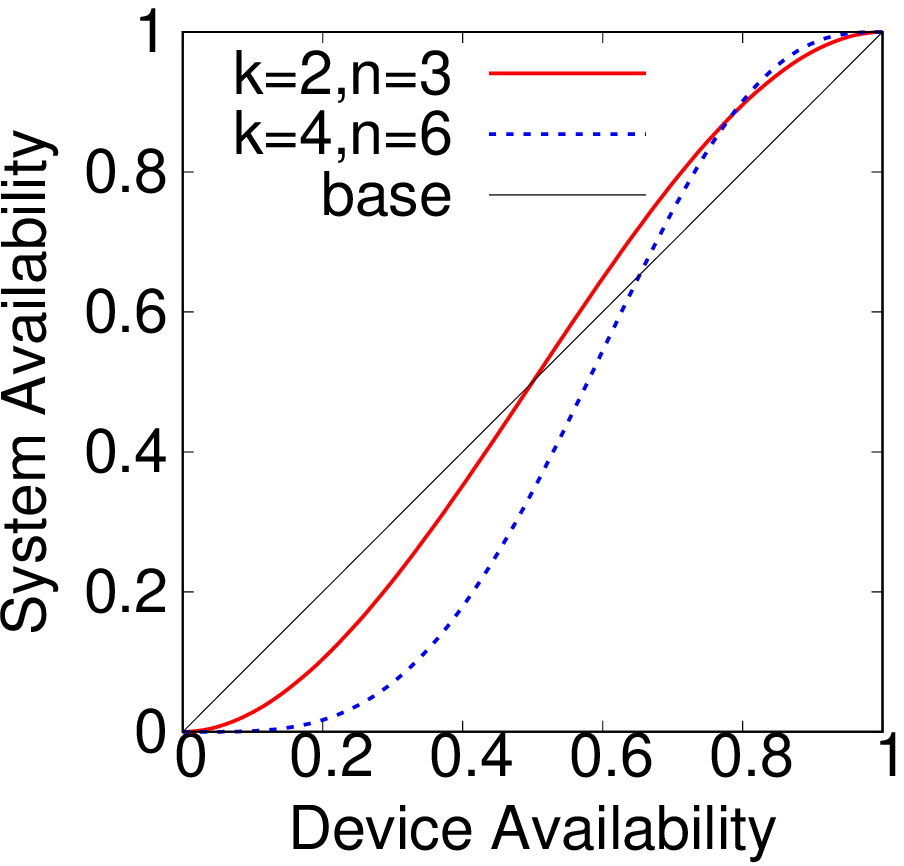}
        \subcaption{k/n=2/3}
        \label{fig:k2n3}
    \end{subfigure}
    \caption{Examples showing how different $(k,n)$ pairs determine different system availability. The example contains three small groups (a, b, c) and each group contains two $(k,n)$ pairs of same ratio. The baseline in each group represents pure local storage}
    \label{fig:devAvail2SysAvail}
\end{figure}

Since both $k$ and $n$ need to be integers, we can easily solve the above minimization problem by iterating over all possible (k,n) pairs and choose those with the minimum costs as solutions. The time complexity of this method is $O(N^2)$. However, there are sometimes several $(k,n)$ pairs with the same minimum costs. To further select among these $(k,n)$ pairs, we need a more precise method to depict the system availability. For simplicity, we assume each device has the same availability $p$. Then, the system availability can be calculated as follows:


\begin{equation}
    \label{eq:compute_sys_avail}
    A(k,n,p)= C^n_kp^k(1-p)^{(n-k)}+ ... +C^{n}_np^{n}
\end{equation}

where $C^n_k$ denotes the number of ways for choosing $k$ from $n$ devices. This equation is complex. In order to get an intuitive understanding of it, we show a few simple examples in Figure~\ref{fig:devAvail2SysAvail}, where we compare the system availability for 6 $(k,n)$ settings calculated based on the above equation. We further divide these six settings into three groups. Within each group, the core rate $k/n$ is identical. As we can see, when the erasure rate increases from $1/3$ to $1/2$ then to $2/3$, the system availability gradually decreases. This indicates the rationality of representing the system availability with $k/n$ for simplicity in Equation (\ref{eq:minimization}). Meanwhile, we observe that, in each group, when the device availability $p$ is small, although the $k/n$ values of different settings are the same, the $(k,n)$ setting with a smaller $n$ has higher availability than the other with a bigger $n$. However, as the device availability $p$ gradually increases over a threshold, the setting with a bigger $n$ starts to achieve higher system availability than the setting with a smaller $n$. Therefore, what $(k,n)$ to choose for a specific $k/n$ is determined by the device availability $p$.

In \schname, for simplicity, we calculate the availability $p_i$ of device $i$ as follows:

\begin{equation}
\label{eq:device_availability}
p_i=\left\{
\begin{aligned}
1 &, & T_i \geq T\\
T_i/T &, & 0<T_i < T 
\end{aligned}
\right.  
\end{equation}
where $T_i$ represents the remaining time of device $i$. When \schname\ selects between $(k_1, n_1)$ and $(k_2, n_2)$ with the same $k/n$ values, it first calculates the value of $A(k_1, n_1, \overline{p})$ and $A(k_2, n_2, \overline{p})$, where $\overline{p}$ represents the average availability of devices, and then chooses the one with a larger value.

\noindent\textbf{Q2: Which specific $n$ devices?} After deciding $(k,n)$, the next question to answer is which $n$ devices to store the $n$ file fragments. \schname\ adopts a simple strategy for this issue. First, it chooses all devices with the remaining storage space larger than $F/k$. Next, it sorts the picked devices based on the expected remaining time in descending order. Finally, it chooses the top $n$ devices with the longest remaining time to store the $n$ file fragments. The complete algorithm for choosing $(k,n)$ and specific $n$ devices is given in Algorithm~\ref{alg:KandN}.

\noindent\textbf{Q3: How are algorithm input parameters decided?} Here we provide a general recommendation for setting $w_a$ and $T$ before data storage tasks are initiated. $w_a$ and $T$ are not meant to be changed for every file; instead, user should set particular values for $w_a$ and $T$ for a particular collection of data. $w_a$ and $T$ values should be set based two factors - how important/mission critical the file is, and how soon user is expected to access/read the data. As an example, for mission critical data such as victim personal image/video files, $w_a$ can be set high such as 0.8, 0.9, 1.0 etc. Also, if user is expected to access the stored data in a near future, user can set an approximate $T$, and the algorithm will choose at least $k$ candidate devices with at least $T$ battery remaining time. Since the algorithm outputs $n$ and $k$, which are integers, fine tuning $w_a$ may not always have impact on the output. Hence, we recommend choosing $w_a$ as a multiple of 0.1.

\subsubsection{Cost Function Lower Bound }
\begin{figure}
	\centering
	\includegraphics[width=0.45\linewidth, angle=-90]{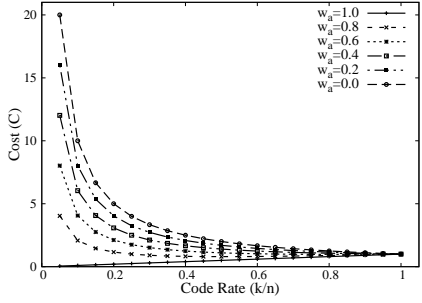}
	\caption{Cost as a function of code rate for different $w_a$}
	\label{fig:cr_vs_cost}
\end{figure}

Figure~\ref{fig:cr_vs_cost} shows how the choice of code rate impacts the cost function for different $w_a$. We identified that for each $w_a$, there is a code rate for which the cost is the lowest, which is the optimal cost. The algorithm tries to reach towards the optimal cost, regardless of the selection of $n$ and $k$ values. For a particular $(n,k)$, if the code rate is similar to the optimal cost code rate, the algorithm will try to hold onto this particular $(n,k)$, unless the devices do not check out storage and battery remaining time requirements (as mentioned in equation~\ref{eq:minimization}). Table~\ref{tab:min_cost} shows the optimal cost for variable $w_a$ and the code rates for which the optimal cost is achieved.

\begin{table}
\centering
\tiny
\begin{tabular}{|c|c|c|c|c|c|c|c|c|}
\hline

$\bm{w_{a}}$ & 1.0 & 0.9 & 0.8 & 0.7 & 0.6 & 0.5 & 0.4 & 0.0 \\
\hline
\textbf{Cost (C)} & 1/N & 0.6 & 0.8 & 0.91 & 0.98 & 1.0 & 1.0 & 1.0 \\
\hline
\textbf{Code Rate} & 1/N & 0.35 & 0.5 & 0.65 & 0.8 & 1.0 & 1.0 & 1.0  \\
\hline
\end{tabular}
\caption{Cost (C) lower bound, as a function of $w_a$ and the corresponding code rate $k/n$ for the lower bound}
\label{tab:min_cost}
\end{table}

A natural question may arise, if the cost for variable $w_a$ is constant, why not use a look-up table to find the code rate with the lowest cost? The answer is, choosing the code rate with the lowest cost does not tell us the exact values of $n$ and $k$ and the particular devices. As an example, for $w_a=0.8$, the code rate 0.5 can be achieved by 15 different combinations of $(n,k)$. So, our algorithm not only chooses code rate with lowest cost, but also chooses devices with minimum required storage and battery remaining time.

\begin{algorithm}[t]
    \SetKwInOut{Input}{Input}
    \SetKwInOut{Output}{Output}
    \Input {$F$, $T$, $S_i$, $T_i$, $w_a$}
    \Output {(k,n) and n devices}
    ${(k,n)} \leftarrow (1,1)$\\
    $C_{min} \leftarrow 1$\\
	\For {$n' \in {1...N}$}
	{
		\For {$k' \in {1...n'}$}
		{
		    \If {Satisfying Eq.(3.2)(3.3)}
		    {
        	    \If {$C(k',n') < C_{min}$}
        	    {
        			${(k,n)} \leftarrow (k',n')$\\
        			$C_{min} \leftarrow C(k',n')$\\
        		}
        		\If {$k'/n' = k/n$}
        	    {
        			\If {$A(k,n, \overline{p}) < A(k',n', \overline{p})$}
        	        {
        	            ${(k,n)} \leftarrow (k',n')$\\
        	        }
        		}
    		}
		}
	}
	$V \leftarrow$ pick up devices with $S_i>F/k$\\
	sort $V$ based on $T_i$ in descending order\\
	$V_n \leftarrow$ choose top $n$ devices with the largest $T_i$\\ 
	return $(k,n)$ and $V_n$
    \caption{Choose $(k,n)$ and $n$ devices}
    \label{alg:KandN}
\end{algorithm}

\subsection{R-Drive Data Retrieval}
\label{subsec:data_ret}

\begin{figure}[t]
	\centering
	\includegraphics[width=0.70\linewidth]{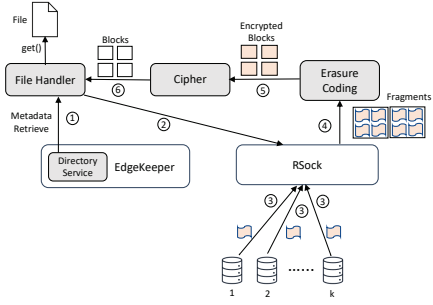}
	\caption{\schname\ file retrieval steps: obtaining from the directory service the location of fragments, deciding which $k$ fragments to retrieve and asking RSock for their delivery, applying erasure coding and re-creating the file from the decrypted blocks}
	\label{fig:distressnet-ng_retrieval}
\end{figure}

Data retrieval in \schname\ involves gathering all blocks of a file and reconstructing it to its original form, as illustrated in Figure~\ref{fig:distressnet-ng_retrieval}. File retrieval is initiated by  calling \textit{get()} API function or executing \textit{-get} command. Directory Service first communicates with EdgeKeeper and fetches the target metadata rnode, given that a rnode for the target file exists and user has permission to access the file. The \textit{fragLocation} field in rnode contains location information of all fragments of all blocks. To reconstruct each block, File Handler must retrieve any $k$ fragments out of $n$, where $k \leq n$. To retrieve any $k$ fragments, File Handler sends fragment requests to $k$ unique devices. Upon receiving a fragment request, a device resolves it by replying with target fragment to the requestor. When $k$ fragment replies are received, File Handler signals Erasure Coding and Cipher components to initiate block decoding and decryption processes respectively. When all the blocks are reconstructed, the original file can also be reconstructed by merging all blocks. All fragment requests and replies are sent/received through RSock.

\subsubsection{Choice of Replica Devices for Data Retrieval}
To choose the replica device to request fragments from, File Handler requests a list of devices from EdgeKeeper with most remaining energy levels and sends $k$ fragment requests to first $k$ devices on the list. In an intermittently connected network environment, a fragment request or reply may be delayed or never be received. One way to deal with this issue is to resend the request for which no reply has been received. The question that still needs to be answered is, how long the sender should wait before initializing resend. \schname\ leverages the TTL in RSock to make sure that a request is resent only when the previous request failed. A fragment requestor can set a TTL within which it wants the reply to be received. If no reply is received within the set time limit, it is guaranteed that the request packet has failed and it is safe to resend the request to a different available replica device.

\subsection{Inter-Edge Data Exchange}

\begin{figure}[t]
\centering
\scalebox{1.10}{
\includegraphics[width=0.90\columnwidth]{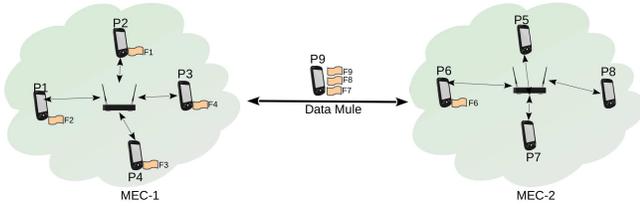}
}
\caption{Inter-edge data storage in R-Drive}
\label{fig:inter_edge}
\end{figure}

This experiment description aims to illustrate inter-edge data storage. Two \distressnetNG edge environments MEC-1 and MEC-2 were set up indoor where both edges had separate WiFi networks and each edge had four phones connected (P1, P2, P3, P4 and P5, P6, P7, P8 respectively). At time T1, P1 stored a 100MB file in \schname\ system with $(N,K)$ values of (8,4) and [P1, P2, P3, P4, P5, P6, P7, P8] as chosen nodes. Hence, P1 aimed to store 8 file fragments [f1, f2, f3, f4, f5, f6, f7, f8], each of 25MB size, to all 8 phones respectively. However, since P1 was not part of MEC-2, some fragments [f5, f6, f7, f8] waited for a link to establish between the two edges. At time T2, a new phone P9 was introduced which toggled WiFi connectivity between two edges and each time stayed connected for approximately 1 minute. The purpose of P9 was to act as a \textit{data mule} between two edges and transfer file fragments from MEC-1 to MEC-2. For first few WiFi toggles, no fragments were transferred to MEC-2. This is due to the fact that, RSock requires a minimum amount of time to learn about the networks and the connectivity pattern. Starting at third WiFi toggle, RSock could successfully transfer the fragments from MEC-1 to MEC-2 and any device in MEC-2 could retrieve the entire file. We repeated the experiment 5 times and the average fragment transfer time was approximately 6 minutes for f5, 8 minutes for f6, and 10 minutes for f7, f8. Figure~\ref{fig:inter_edge} shows the inter-edge data storage in R-Drive.

\section{R-Drive Implementation}

\begin{figure}[t]
\centering
\includegraphics[width=0.70\linewidth]{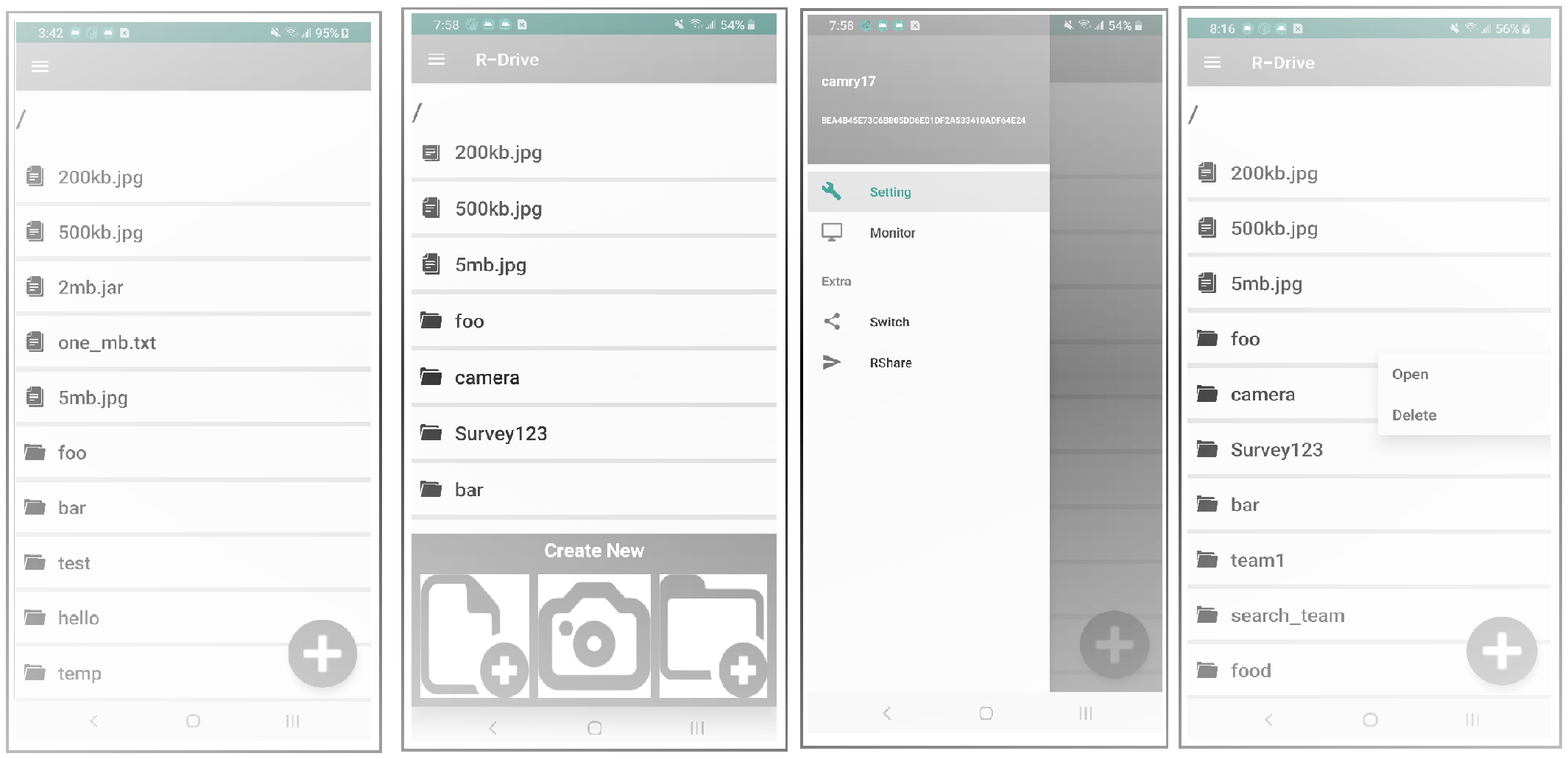}
\caption{\schname\ Android Application}
\label{fig:android_app}
\end{figure}

We implemented \schname\ as an app for Android mobile devices and as a daemon process for Linux desktops (HPC nodes). The implementation of \schname\ has approximately 10,000 lines of Java code. The Android app (shown in Figure~\ref{fig:android_app}) is compatible with Android version 7.1, 8.0, 10.0. The app runs as an Android background service aiming for hands-free use. \schname\ home page shows the current directory name and the files and folders in the current directory. Long pressing on each item will open a floating menu that allows a user to either open or delete the file/folder. On the top left, the hamburger icon opens a navigation drawer where users can find options to change application setting.
The File Handler module (shown in Figure~\ref{fig:r-drive_architecture}) exposes the \schname\ Java API and also handles buttons from the Android app. The module uses the \textit{javax.crypto} package for the 256 bit AES encryption, as the Cipher. For Shamir's Secret Sharing algorithm, \schname\ employs \textit{secretsharejava}~\cite{secretsharejava}, an open source library implementing the LaGrange Interpolating Polynomial Scheme~\cite{schneier2007applied}. We used BackBlaze~\cite{backblaze} Reed-Solomon erasure coding library, an open source implementation available for both academic and commercial use.

\schname\ provides command line interface (CLI) for Linux desktop users, to perform storage operations on remote devices if the device operators allow permissions. \schname\ commands are interpreted by the Command Handler. Command Handler consists of a hand-written lexer and parser. Lexer takes an input command as text stream, converts into a series of tokens and parser converts the tokens into a parse-tree. The parse-tree enables Command Handler to identify the type of command. Below is the grammar \schname\ uses for file system commands.

\begin{small}
\begin{verbatim}
COMMAND::= 'dfs' OPTION ARGUMENT
OPTION::= -put | -get | -mkdir | -ls | -rm 
    | -setfacl | -getfacl
ARGUMENT::= PATH | PERMISSION | PATH PERMISSION
PATH::= <local_path> | <rdrive_path> 
    | <local_path>  <rdrive_path>
PERMISSION::= 'OWNER' | 'WORLD' | USERS
USERS::= GUID | USERS GUID
GUID::= <40 bytes ASCII printable characters>
\end{verbatim}
\end{small}

Here \textit{local\_path} means the local absolute path of a file in local file system. \textit{rdrive\_path} means either a file or a directory path in \schname\ file system. GUID is a unique 40 bytes long string comprising both characters and numbers.

\section{R-Drive Performance Evaluation}
\label{sec:evaluation}
We employed two systems for \schname\ benchmarking: 1) NIST Public Safety Communications Research (PSCR) deployable system, equipped with LTE (Star Solutions COMPAC-N) and Wi-Fi (Ubiquiti EdgerouterX) networks. The system has 10MHz downlink and uplink channels, the observed LTE data rates are about 95 and 20Mbps respectively. 2) \distressnetNG manpack system, which can be carried in a backpack of a first responder. It also consists of both LTE (BaiCells Nova 227 eNB) and Wi-Fi (Unify 802.11AC Mesh) networks, and Intel Next Unit of Computing Kit (NUC) as application server. For 20MHz downlink and uplink channels, the LTE can provide a maximum data rate of 110 and 20Mbps respectively~\cite{baicells}. Wi-Fi are capable of providing around 100Mbps data transfer rate. We used 15 Android devices of Essential PH-1, Samsung S8 and Sonim XP8 devices with Android versions 7.1, 8.0 and 10.0.  

\subsection{Adaptive Erasure Coding Evaluation}
In this section, we provide an in-depth analysis of how $w_a$ parameter impacts the choice $(k, n)$ values, hence also the code-rate and $F'$(file size after erasure coding). We also analyze the choice of code-rate and it's impact on the cost function. We performed experiments on variable network sizes (10, 20, 30), for a file size $F$ of 500MB, and expected file availability time $T$ of 300 minutes. The storage $S_i$ and expected battery remaining times $T_i$ for nodes were generated using pseudo-random value generator with mean-variance of $(100, 20)$ and $(300, 80)$ respectively. The experiments were conducted for 30 runs before results were averaged.

\subsubsection{Achieved cost for variable $w_a$}
Table~\ref{tab:achieved_cost} shows the average achieved cost for variable $w_a$ and network size. With larger network size the cost function is computed over more combinations of $(n,k)$ values, hence the algorithm achieves cost value closer to optimal value.

\begin{table}[!htb]
\centering
\scalebox{0.85}{
\begin{tabular}{|c|c|c|c|c|}
    \cline{3-5}
    \hhline{|-----|} \multicolumn{1}{|c|}{$\bm{w_{a}}$} & \multicolumn{1}{c|}{\textbf{Lower}} & \multicolumn{3}{c|}{\textbf{Achieved Cost}}\\
    \hhline{~~---}
    \multicolumn{1}{|c |}{} & \multicolumn{1}{c|}{\textbf{Bound}} & NS=30 & NS=20 & NS=10\\
    \hhline{|-----}
    1.0 & 0.00 & 0.2402 & 0.3613 & 0.66 \\
    \hhline{|-----|}
    0.9 & 0.6 & 0.6 & 0.6048 & 0.6782 \\
    \hhline{|-----|}
    0.8 & 0.8 & 0.8 & 0.8121 & 0.8360 \\
    \hhline{|-----|}
    0.7 & 0.9165 & 0.9165 & 0.9166 & 0.9183 \\
    \hhline{|-----|}
    0.6 & 0.9797 & 0.9797 & 0.9799 & 0.9807 \\
    \hhline{|-----|}
    0.5 & 1.0 & 1.0 & 1.0 & 1.0 \\
    \hhline{|-----|}
\end{tabular}
}
\captionsetup{belowskip=7pt}
\caption{Achieved cost for variable $w_a$ and Network Size (NS)}
\label{tab:achieved_cost}
\end{table}

\subsubsection{Impact of $w_{a}$ and Network Size on Code Rate and $F'$}


\begin{figure}[h!]
    \centering
\begin{subfigure}{.8\columnwidth}
\centering
    \includegraphics[height=.7\columnwidth, angle=-90]{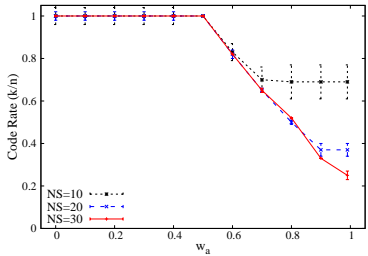}
    \subcaption{}
    \label{fig:wa_vs_cr}
\end{subfigure}

\begin{subfigure}{.8\columnwidth}
    \centering
        \includegraphics[height=0.7\columnwidth, angle=-90]{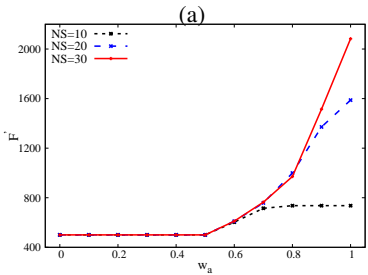}
        \subcaption{}
        \label{fig:wa_vs_fstr}
\end{subfigure}
    \caption{Effect of $w_a$ on: a) code Rate ($k/n$); and b) file size $F'$, for network sizes, NS=10, 20 and 30}
    \label{fig:wa_vs_cr_vs_fStr}
\end{figure}

Figure~\ref{fig:wa_vs_cr} illustrates that with increased $w_a$, the code rate decreases. This is expected, since the algorithm takes $w_a$ as an input for the weight of availability. With larger $w_a$, the algorithm chooses a larger $n$ in an attempt to provide more data redundancy, hence the code rate decreases. Figure~\ref{fig:wa_vs_fstr} shows the $F'$ as a function of $w_a$. $F'$ increases exponentially with higher $w_a$. Since chosen code rate is higher for network size 10, $F'$ is higher compared to network size 20 and 30.

\begin{figure}[h!]
\centering
\begin{subfigure}{.9\linewidth}
\centering
    \includegraphics[width=0.65\linewidth]{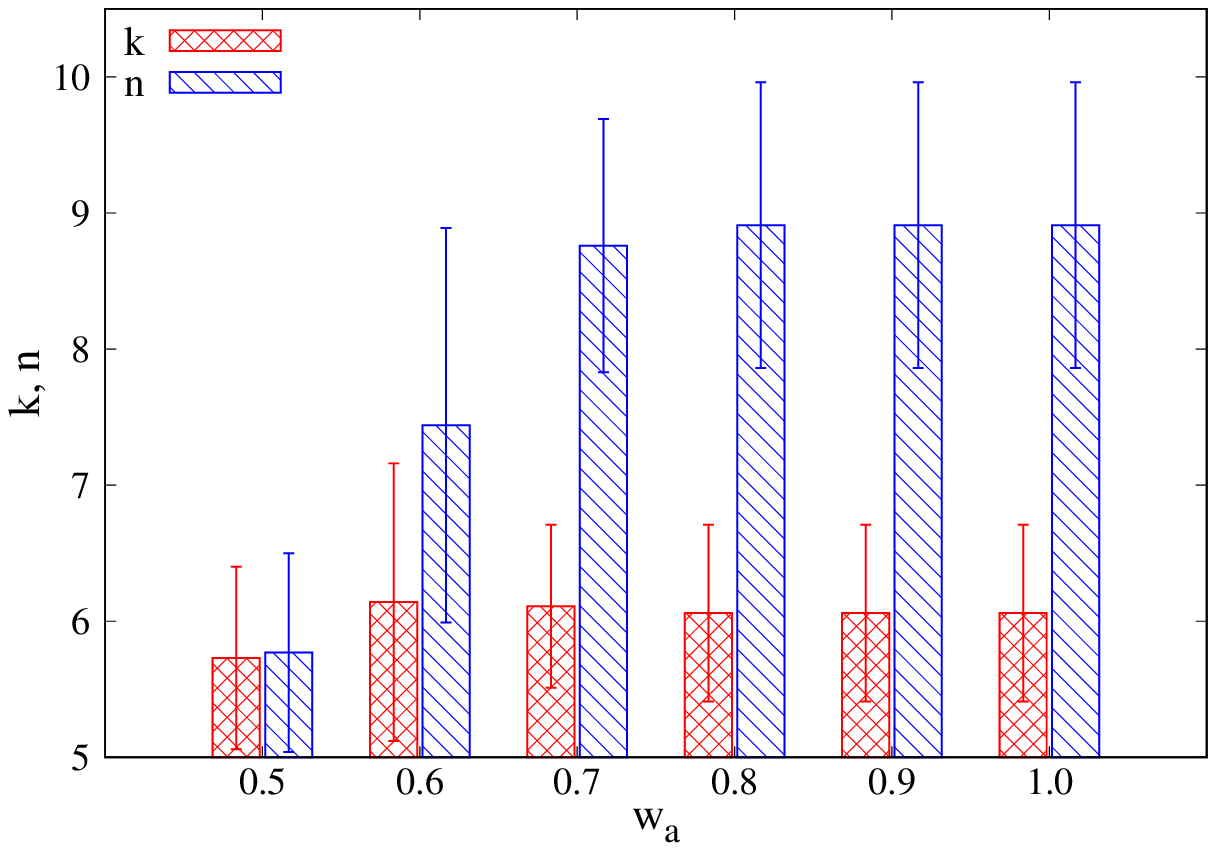}
    \subcaption{}
    \label{fig:ave_nk_10}
\end{subfigure}
\begin{subfigure}{.9\linewidth}
    \centering
    \includegraphics[width=.65\linewidth]{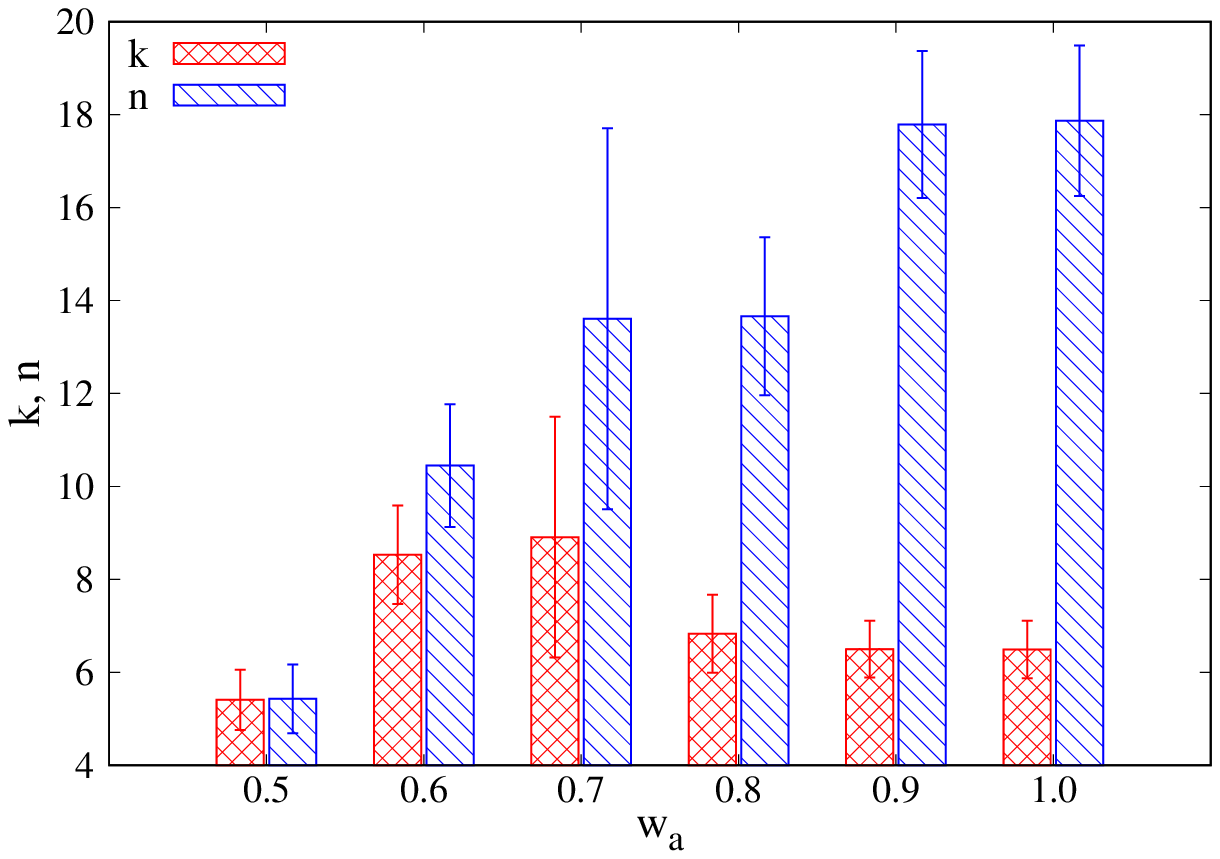}
    \subcaption{}
    \label{fig:ave_nk_20}
\end{subfigure}
\begin{subfigure}{.9\linewidth}
    \centering
    \includegraphics[width=.65\linewidth]{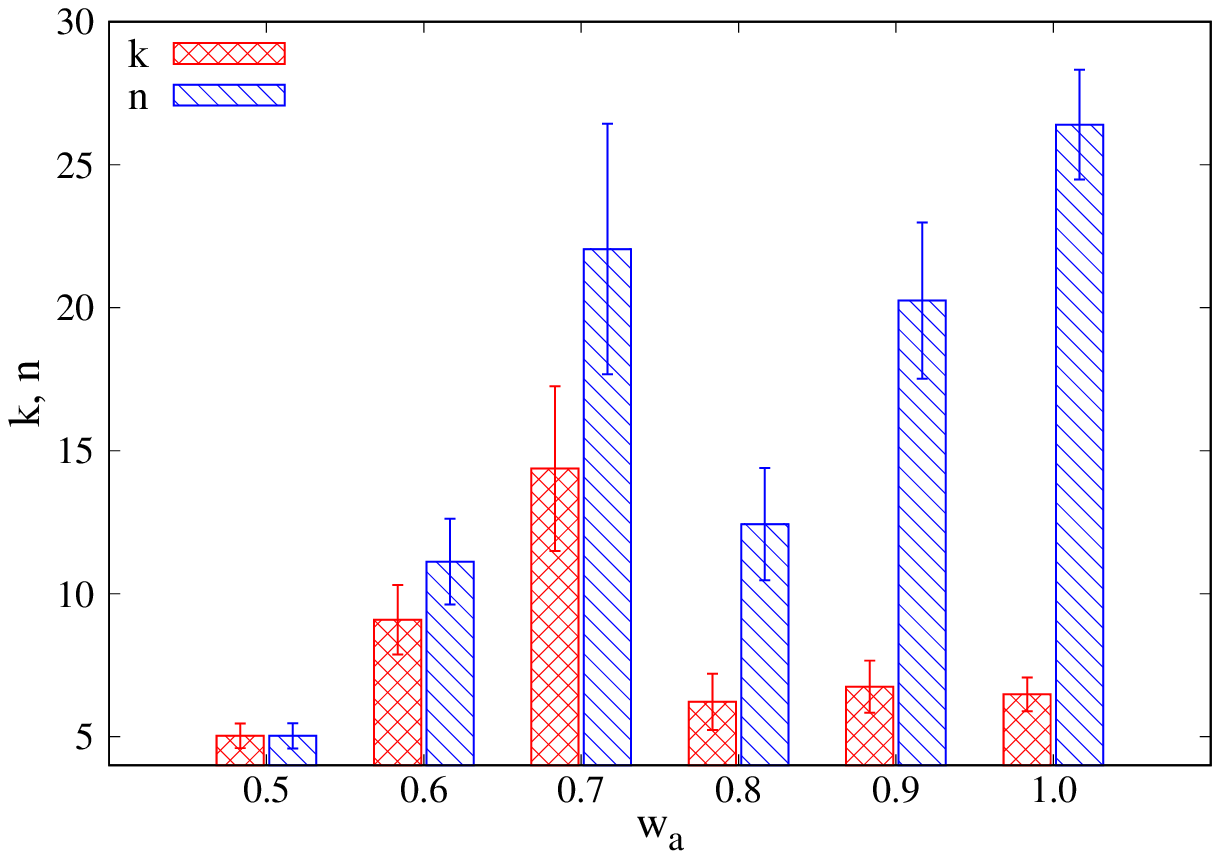}
    \subcaption{}
    \label{fig:ave_nk_30}
\end{subfigure}
\caption{Average $(k,n)$ for different network sizes NS: a) 10; b) 20; and c) 30}
\label{fig:ave_nk}
\end{figure}

Figure~\ref{fig:ave_nk} shows the averages of chosen $n$ and $k$ values for variable network size over 30 iterations. For higher $w_a$, the algorithm chooses larger $n$ value to provide data redundancy. In Figure~\ref{fig:ave_nk_30}, for $w_a$ 0.8, the chosen $(n,k)$ values are lower than the values selected for 0.7. This is because for $w_a$ of 0.8, the optimal cost code rate is 0.5, and the algorithm produced resultant $(n,k)$ values of (10,5), (12,6), (14,7) over 30 runs that averaged to (13.07, 6.53).





\subsection{Directory Service Resilience}
\begin{figure}[ht]
\centering
\includegraphics[width=0.9\columnwidth]{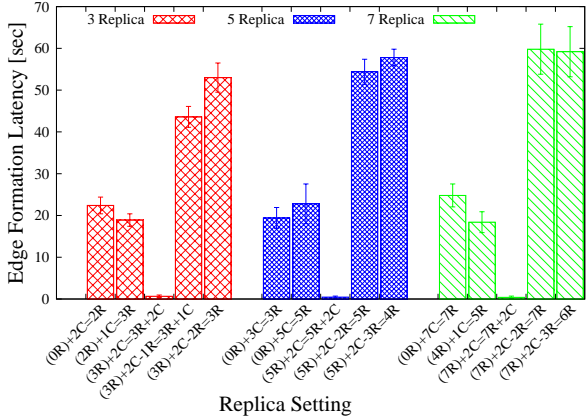}
\caption{Edge formation latency for variable EdgeKeeper replica settings. Here \textbf{R} and \textbf{C} denote for replica and client respectively}
	\label{fig:edge_reform_3_5_7}
\end{figure}

We measured resilience of \schname\ Directory Service based on how fast Directory Service becomes operational after failure event takes place due to several reasons such as configuration changes, device or network failures etc. EdgeKeeper reforms a new ensemble with available devices as soon as it detects that previous one is broken. The experiment was conducted at NIST testbed with Samsung S8 phones and LTE networking backbone. Each bar in Figure~\ref{fig:edge_reform_3_5_7} shows average delay of reforming an edge after an event takes place. For each x-axis ticks, the equation describes the event. The term inside parentheses on the left side of the equation represents an initial condition and the remaining terms represents the changes that have been introduced. The right side of the equation represents the final stable condition. As example, \textbf{(3R)+2C-1R=3R+1C} denotes that, in an stable ensemble of 3 devices, we simultaneously added 2 new devices and took out 1 replica device, and measured how long it took to reform the ensemble with 3 devices. When an ensemble is stable, only adding new devices takes very small amount of time, as the new devices only join as clients and no reformation takes place.

\begin{figure*}
\begin{subfigure}{\columnwidth}
  \centering
  \includegraphics[width=.95\columnwidth]{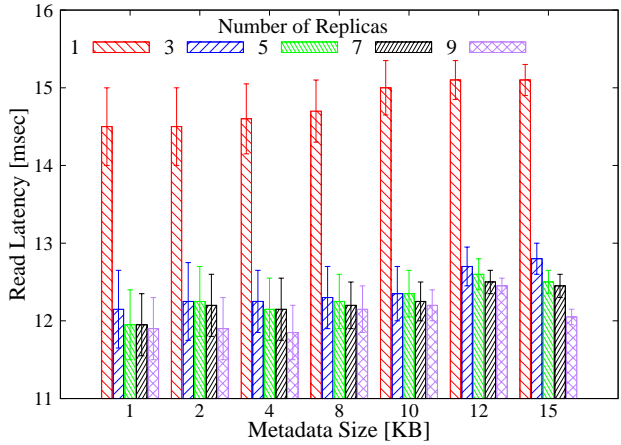}
  \caption{}
  \label{fig:get_throughput}
\end{subfigure}
\begin{subfigure}{\columnwidth}
  \centering
  \includegraphics[width=.95\columnwidth]{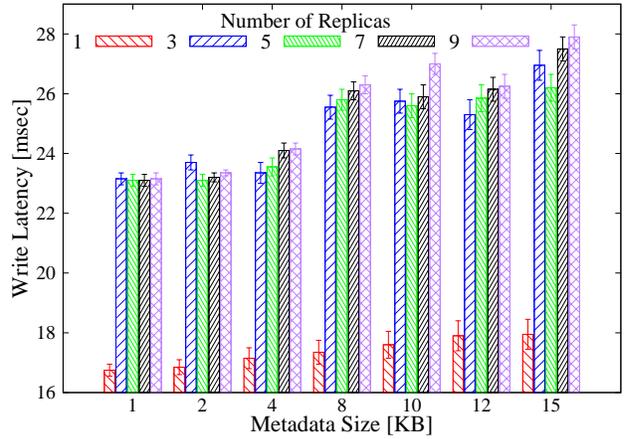}
  \caption{}
  \label{fig:put_throughput}
\end{subfigure}
\caption{Metadata read (a) and write (b) latencies as a function of metadata size, for link availability 1.0}
\label{fig:metadata_throughput_1_0}
\end{figure*}


\begin{figure*}
\begin{subfigure}{\columnwidth}
  \centering
  \includegraphics[width=.95\columnwidth]{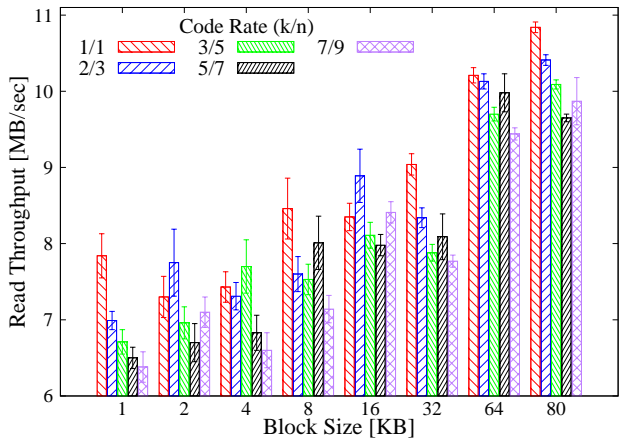}
  \caption{}
  \label{fig:data_read_throughput_0_5}
\end{subfigure}
\begin{subfigure}{\columnwidth}
  \centering
  \includegraphics[width=.95\columnwidth]{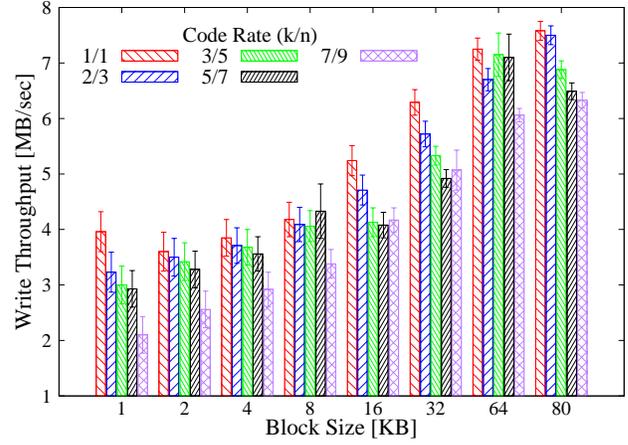}
  \caption{}
  \label{fig:data_write_throughput_0_5}
\end{subfigure}
\begin{subfigure}{\columnwidth}
  \centering
  \includegraphics[width=.95\columnwidth]{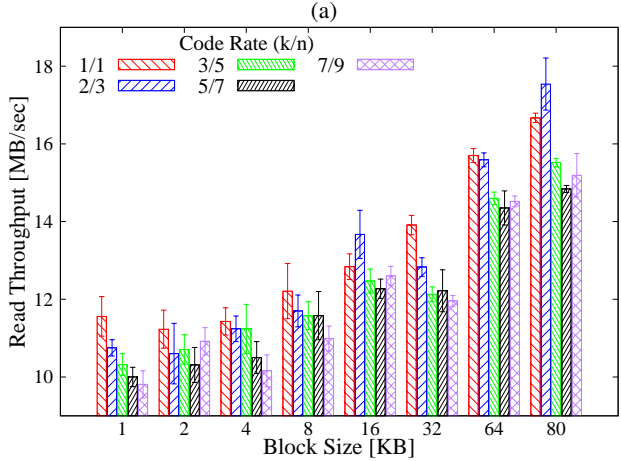}
  \caption{}
  \label{fig:data_read_throughput_1_0}
\end{subfigure}
\begin{subfigure}{\columnwidth}
  \centering
  \includegraphics[width=.95\columnwidth]{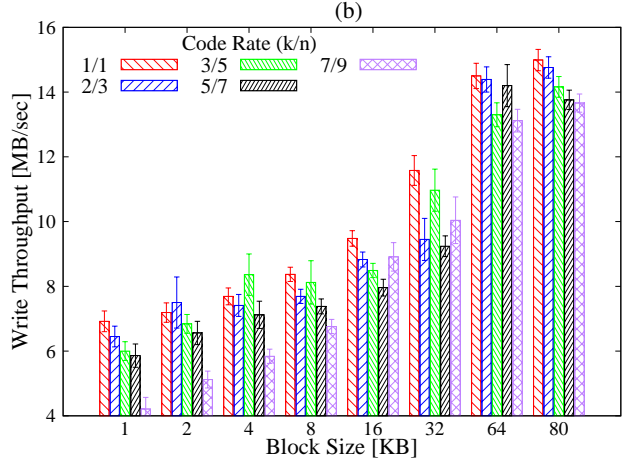}
  \caption{}
  \label{fig:data_write_throughput_1_0}
\end{subfigure}
\caption{Data read and write throughput, for 0.5 link availability (a, b) and 1.0 link availability (c, d)}
\label{fig:data_throughput_1_0}
\end{figure*}

\subsection{Directory Service Latency}
 Figure~{\ref{fig:get_throughput}} and ~{\ref{fig:put_throughput}} show the average metadata read and write latencies on 9 Android phones, for variable metadata sizes and number of EdgeKeeper servers, using both Wi-Fi and LTE. Figure~{\ref{fig:get_throughput}} shows that for each metadata size group, as the number of EdgeKeeper servers increases more than 1, metadata retrieval latency drops. This is because more servers can perform better load balancing, resulting in overall lower retrieval latency. Variable metadata sizes have very little effect on retrieval latency. As the range of metadata size is very small, usually within 1 to 15KB, the average cost to fetch most metadata is almost the same. Figure~{\ref{fig:put_throughput}} also shows that, as number of servers increases more than 1, write latency increases significantly, due to the fact that, having more than 1 server brings additional cost to check for quorum among servers before the data is committed. For both read and write, adding more servers does provide additional fault tolerance, but does not minimize latency significantly. 



\subsection{Data Throughput}

Figures~\ref{fig:data_throughput_1_0} show the average data read and write throughput for variable code rates, block sizes and link availability. The experiments were conducted with 9 Android devices with pure Wi-Fi and LTE connectivity. Each phone stored and retrieved 3GB of data simultaneously, comprising of file sizes ranging between 10 to 200MB. We controlled the link availability using another android application that can turn on/off networking based on a presetting probability. As figures suggest, read/write throughput is higher in a purely connected network compared to loosely connected network. Also, increasing block size increases throughput for both read and write due to higher block size ensures lower block count, resulting in lower number of total fragments to transfer over network. Moreover, for most block size groups, throughput slightly drops with lower code rates due to lower code rate comes with higher n and k values, resulting in more fragments to be distributed or retrieved respectively.

We compared data read/write throughput of \schname\ with MDFS~\cite{DBLP:journals/tcc/ChenWSX15}, which resembles the closest with \schname\ in terms of design paradigm. For 2MB files and $(n,k)$ parameters as $(7,3)$, MDFS provided 2.3MB/sec and 2.0MB/sec of read and write throughput respectively, whereas \schname\ provides 11.5 and 6.5 MB/sec for read and write respectively.






\subsection{R-Drive Overhead}


\subsubsection{Memory Footprint}
We traced real-time memory footprint for \schname\ using Android Profiler~\cite{androidprofiler} during file storage and retrieval process. Table~\ref{tab:object_alloc_dealloc} shows the average heap object allocation and deallocation during file creation and file retrieval for variable iterations. The number of dangling objects starts to increase over time as number of file creation/retrieval increases.

\begin{table}[ht]
        \small
		\centering
		\captionsetup{belowskip=7pt}
		\scalebox{0.85}{
		\begin{tabular}{|c|c|c|c|c|}
		\hline
		\textbf{Count} & \multicolumn{2}{c|}{File Creation} & \multicolumn{2}{c|}{File Retrieval}\\
         \cline{2-3}  \cline{4-5}
		& \textbf{Alloc} & \textbf{Dealloc} & \textbf{Alloc} & \textbf{Dealloc} \\
		\hline
		10 & 4,381 & 4,381 & 2,526 & 2,526 \\
		\hline
		100 & 43,885 & 43,876 & 25,298 & 25,288 \\
        \hline
        1,000 & 438,911 & 438,884 & 253,012 & 252,996 \\
        \hline
        \end{tabular}
		}
		\caption{Number of allocated and de-allocated objects for different numbers of file creation and retrieval}
		\label{tab:object_alloc_dealloc}
\end{table}


\subsubsection{Energy Consumption}
\begin{table}[ht]
		\centering
		\captionsetup{belowskip=7pt}
		\scalebox{0.75}{
		\begin{tabular}{|c|c|c|c|c|c|}
		\hline
		\textbf{Device} & \textbf{Runtime} & \multicolumn{3}{c|}{\textbf{Consumed}} & \textbf{Dist-NG} \\
		\cline{3-5} 
		\textbf{} &  \textbf{h:min} & \textbf{\%} & \textbf{mAh} & \textbf{Wh} & \textbf{Wh} \\
		\hline
		[1] & 3:30 & 12.5 & 377.4 & 1.5 & 3.5\\
		\hline
		[2] & 3:05 & 11.9 & 323.5 & 1.2 & 3.2\\
        \hline
        [3]  & 3:15 & 12.6 & 381.8 & 1.5 & 3.8\\
        \hline
        \end{tabular}
        }
		\caption{\schname\ energy consumption for different Android devices: Samsung S8 [1]; Google Pixel 2 [2]; and Essential PH1 [3]}
		\label{tab:battery_cons}
\end{table}

Table~\ref{tab:battery_cons} shows \schname\ application average energy consumption for continuous workload in different Android devices. We used Battery Historian~\cite{batteryhistorian} to pull Android battery usage data for 100\% to 0\% exhaustion. We deduce that, if similar devices are used in field, first responders may need to switch the device battery after approximately 3.5 hours. 

\subsubsection{Processing Overhead}
We measured processing time for components responsible for encryption key generation, data encryption and data erasure coding, as shown in Table~\ref{tab:processing_overhead}. We observed that, data encryption takes the majority amount of processing time.

\begin{table}[ht]
        \small
		\centering
		\captionsetup{belowskip=7pt}
		\scalebox{1.10}{
		\begin{tabular}{|c|c|c|c|}
		\hline
		\textbf{} & \textbf{Shamir} & \textbf{AES} & \textbf{Reed-Solomon} \\
		\hline
		Read & 5\% & 87\% & 8\% \\
		\hline
		Write & 3\% & 84\% & 13\% \\
        \hline
        \end{tabular}
		}
		\caption{Processing overhead as percentage of the total delay}
		\label{tab:processing_overhead}
\end{table}

\subsubsection{Algorithm Execution Time}
Table~\ref{tab:alg_execution_time} shows the average algorithm execution time in milliseconds on a Samsung S8 Android device of average over 1000 iterations.

\begin{table}[ht]
        \small
		\centering
		\captionsetup{belowskip=7pt}
		\scalebox{1.10}{
		\begin{tabular}{|c|c|c|c|}
		\hline
		\textbf{Device} & \textbf{NS=30} & \textbf{NS=20} & \textbf{NS=10} \\
		\hline
		Samsung S8 & 101.6msec & 15.3msec & 0.5msec \\
		\hline
        \end{tabular}
		}
		\caption{Execution time of the Adaptive Coding algorithm in Samsung S8}
		\label{tab:alg_execution_time}
\end{table}

\section{Conclusions and Future Work}
\vspace{10pt}
\label{sec:conclusions}
We presented R-Drive, a device and network failure resilient data storage and sharing framework, with detailed explanation of how R-Drive resiliently store and share data leveraging edge devices in an environment where network connectivity constantly fluctuates. We presented and evaluated the implementation for various parameters such as device availability, network availability, block sizes, code rates etc. For future work, we want to investigate how to reduce data encryption delay using more robust library. We also plan to investigate fragments transfer from one vulnerable device to a safe one over opportunistic network before device failure.

\bibliographystyle{acm}
\bibliography{references}


\end{document}